\documentclass[11pt, a4paper]{article}
\usepackage[utf8]{inputenc}
\usepackage{amsmath}
\usepackage{amssymb}
\usepackage{geometry}
\usepackage{hyperref}
\usepackage{booktabs}

\title{\textbf{A Topos-Theoretic Interpretation of Blockchain Systems:} \\ Sheaves of Consensus and the Logic of Decentralized Truth}
\author{\textbf{Manuel Hernández}, \textbf{Eduardo Sánchez-Soto}}
\date{December 12, 2025}

\begin{document}

\maketitle

\begin{abstract}
The predominant formal models for blockchain systems, particularly smart contracts, have largely been drawn from the classical theory of computation, with the finite state machine (FSM) or labeled transition system serving as the primary conceptual tool. However, the FSM relegates the most difficult and novel aspect of a blockchain---the achievement of consensus in a decentralized environment---to a complex, often messy, implementation detail that lies outside the formal model itself. But the process of consensus is not an ancillary feature; it is the very essence of the computational phenomenon. To model it faithfully, a new mathematical language is required. The central thesis of this work is that topos theory, the theory of categories of sheaves, provides the native mathematical language for systems defined by local consistency and the construction of global truth.
\end{abstract}

\section{Introduction: Beyond State Machines}
\subsection{Motivation}
The predominant formal models for blockchain systems, particularly smart contracts, have largely been drawn from the classical theory of computation, with the finite state machine (FSM) or labeled transition system serving as the primary conceptual tool \cite{lambert_topos_blockchain}. While such models are effective in describing the deterministic execution of a single contract's bytecode on an already-agreed-upon state, they are fundamentally inadequate for capturing the defining characteristic of the system as a whole: the distributed, asynchronous, and adversarial process of \emph{constructing} a single, canonical state from a multitude of partial, potentially conflicting local views.

The FSM model presupposes a singular, objective reality---a global state---and concerns itself only with the rules for transitioning from one such state to the next. It relegates the most difficult and novel aspect of a blockchain---the achievement of consensus in a decentralized environment---to a complex, often messy, implementation detail that lies outside the formal model itself. This report posits that this is a profound categorical error. The process of consensus is not an ancillary feature; it is the very essence of the computational phenomenon. To model it faithfully, a new mathematical language is required. The central thesis of this work is that topos theory, the theory of categories of sheaves, provides the native mathematical language for systems defined by local consistency and the construction of global truth \cite{maclane_moerdijk_sheaves}.

\subsection{The Philosophical Imperative for a New Model}
\label{subsec:philosophical}
The inadequacy of classical models for distributed systems mirrors a long-standing philosophical critique of classical logic and set theory. In his work on the logic of sheaves, Xavier Caicedo argues that classical models often rely on ``limit-idealizations''---abstract, point-like entities that have no direct correlate in observable reality \cite{caicedo_limit, caicedo_sheaves}. An assertion about a point in a topological space, for instance, is truly an assertion about its properties over an open neighborhood; the point itself is an ideal limit. A blockchain's global state at a specific block height is precisely such a limit-idealization. The true object of study, the empirically accessible phenomenon, is the extended, evolving process of state validation and agreement that unfolds over the history of the chain and across the network's topology.

This reframes the central challenge of distributed computing, the ``Problem of Global State,'' as a specific instance of the classical philosophical ``Problem of Universals'' \cite{caicedo_limit}. In this parallel, classical computational models, much like nominalist philosophies, struggle to account for the universal property of a global state without postulating an abstract, centralized entity---a ``master observer'' or a ``canonical state-holder.'' This is an entity that a blockchain is explicitly designed to eliminate. Nominalism attempts to avoid commitment to abstract universals, yet often finds itself unable to describe reality without them. Similarly, classical distributed systems models often abstract away the messiness of consensus by assuming its eventual success, thereby implicitly invoking a centralizing force that the protocol itself is meant to construct. The challenge, then, is to find a logical and mathematical framework that can describe a universal property (global state) as emerging directly from a collection of particulars (local views) without postulating its \emph{a priori} existence.

\subsection{Topos Theory as the Framework for Contextual Reality}
Topos theory, emerging from algebraic geometry and mathematical logic, provides exactly such a framework \cite{maclane_moerdijk_sheaves, goldblatt_topoi}. A topos is, most fundamentally, a category that behaves like the category of sets, but where the notion of ``set'' is generalized to an object that can vary over a ``base space'' of contexts. In a topos of sheaves, this structure is made explicit. It is a category of sheaves on a site, which consists of a base category of contexts and a Grothendieck topology defining what it means for a collection of contexts to ``cover'' another \cite{maclane_moerdijk_sheaves}. This framework is the natural setting for modeling phenomena that are context-dependent and locally determined.

This report constructs a model where the base space represents the possible histories of a blockchain. Data that varies over these histories, such as the set of unconfirmed transactions, will be modeled as a \emph{presheaf}. A consistency condition for this data, which ensures that local information can be ``glued'' together into a coherent whole, is the \emph{sheaf axiom}. This axiom is the mathematical formalization of a consensus algorithm. The resulting topos possesses an internal logic that is inherently contextual and, in general, intuitionistic \cite{goldblatt_topoi, fourman_scott_logic}. This logical structure treats the ``execution ambiguity'' inherent in distributed systems not as a problem to be abstracted away, but as the fundamental geometric structure of the computational space itself \cite{lambert_topos_blockchain}.

The adoption of this framework represents a necessary philosophical shift in how distributed computation is understood. It moves the analytical focus from the static state of the system to the dynamic process of observation and agreement about the state. The truth of a proposition within the system is not a pre-existing Platonic fact but is something that is actively constructed and can only be known through a consistent fabric of local observations---a perspective that aligns deeply with the constructive philosophy of intuitionistic mathematics.

\section{The Site of Computation: Defining the Spacetime of a Blockchain}
To construct a topos of sheaves, one must first define the site upon which the sheaves will live. A site is a pair $(C, J)$, consisting of a small category $C$ and a Grothendieck topology $J$ on $C$ \cite{maclane_moerdijk_sheaves}. This site forms the foundational structure of our model, representing the ``computational spacetime'' of all possible blockchain evolutions. It defines the universe of possible states and the causal relationships between them, as well as the rules governing liveness and progress.

\subsection{The Category of Histories ($C$)}
The base category of our site, denoted $C$, is the category of valid partial histories. This construction formalizes the append-only, causal structure of a blockchain ledger and is analogous to the protocol complexes used in topological models of distributed computing, which represent the set of all possible execution prefixes \cite{herlihy_shavit_multiprocessor, fajstrup_components}.
\begin{itemize}
    \item \textbf{Objects:} An object $h$ in $C$ is a \emph{valid partial history} of the blockchain. Formally, this is a finite, ordered sequence of valid blocks $(b_0, b_1, \dots, b_n)$ where $b_0$ is the designated genesis block. Each block $b_k$ for $k > 0$ must be valid according to the protocol rules, which typically includes containing a valid cryptographic hash of the preceding block $b_{k-1}$, a valid proof-of-work (or other consensus-related data), and a set of valid transactions. An object in $C$ thus represents a specific, observable state of the ledger at a finite point in time.
    \item \textbf{Morphisms:} A morphism $f: h \to h'$ in $C$ exists if and only if the history $h'$ is a valid extension of the history $h$. That is, $h = (b_0, \dots, b_n)$ and $h' = (b_0, \dots, b_n, b_{n+1}, \dots, b_m)$ for some $m \geq n$. The existence of a morphism from $h$ to $h'$ signifies that $h'$ is a possible future state of a system that has observed history $h$.
    \begin{itemize}
        \item \textbf{Composition:} If there are morphisms $f: h \to h'$ and $g: h' \to h''$, their composition $g \circ f: h \to h''$ is the morphism corresponding to the extension from $h$ to $h''$. Since extensions are transitive, this composition is always well-defined.
        \item \textbf{Identity:} For any object $h$, the identity morphism $\text{id}_h: h \to h$ is the empty extension, where $h'$ is simply $h$.
    \end{itemize}
\end{itemize}
This categorical structure precisely captures the causal, append-only nature of the blockchain. The objects are snapshots of the system's evolution, and the morphisms represent the relentless, forward-only progression of time and computation.

\subsection{The Grothendieck Topology of Liveness ($J$)}
A Grothendieck topology $J$ on $C$ endows the category with a notion of ``covering'' or ``locality'' \cite{maclane_moerdijk_sheaves}. It is a function that assigns to each object $h \in C$ a collection of \emph{covering sieves} $J(h)$. A sieve $S$ on $h$ is a collection of morphisms with codomain $h$ (in the opposite category $C^{\text{op}}$, which corresponds to morphisms with domain $h$ in $C$) that is closed under pre-composition. Intuitively, a covering sieve represents a collection of ``local pieces'' that are jointly sufficient to determine properties at $h$.

In our model, we define the topology $J$ as the \emph{Grothendieck Topology of Liveness}. A sieve on an object $h$ is considered a covering sieve if it represents a set of possible future developments such that the underlying consensus protocol guarantees that at least one of them will eventually occur, assuming the system remains ``live'' and is not subject to catastrophic failure. This directly encodes the liveness and fault-tolerance assumptions of the protocol into the geometry of the site \cite{lynch_distributed}.
\begin{itemize}
    \item \textbf{Definition:} A sieve $S$ on a history $h$ is a \emph{liveness cover} if, under the assumptions of the consensus protocol, the system is guaranteed to eventually transition to a state $h'$ such that the morphism $h \to h'$ is in $S$.
    \item \textbf{Example (Proof-of-Work):} For a history $h$, consider the set of all morphisms $f_i: h \to h_i$ where each $h_i$ is $h$ extended by a single, distinct, valid block $b_i$. The sieve generated by this family of morphisms consists of all possible valid one-block extensions of $h$ and all further extensions thereof. This sieve is a liveness cover. The core liveness assumption of a Proof-of-Work protocol is that, as long as honest hashing power is active on the network, \emph{some} miner will eventually succeed in finding and propagating a valid next block. The topology does not specify \emph{which} extension will occur, only that the collection of all valid immediate possibilities constitutes a cover. A fork occurs when multiple such extensions are found and propagated concurrently, creating multiple paths into the future, all of which are part of the covering sieve.
    \item \textbf{Example (Byzantine Fault Tolerance):} In a BFT-style protocol with a fixed set of validators, a covering sieve for a history $h$ might be defined differently. A cover could be the sieve generated by all extensions $h \to h'$ where the new block in $h'$ has been signed by a quorum (e.g., $2f+1$) of validators. The liveness assumption here is that a sufficient number of validators are honest and connected, and will eventually agree upon and sign a next block.
\end{itemize}
This demonstrates a deep structural connection: the choice of the Grothendieck topology is the formal specification of the consensus protocol's threat model and liveness guarantees. Different consensus mechanisms (e.g., Nakamoto consensus vs. Tendermint) correspond to different Grothendieck topologies on the same underlying category of histories $C$. The site $(C, J)$ is thus a unified model of both the data structure (the chain, represented by $C$) and the protocol that governs its growth (the consensus rules, represented by $J$).

To ground these abstract definitions, Table~\ref{tab:mapping} provides a direct mapping between the core concepts of the topos-theoretic model and their concrete interpretations in a blockchain system.

\begin{table}[h!]
\centering
\small
\begin{tabular}{ll}
\toprule
\textbf{Topos-Theoretic Concept} & \textbf{Proposed Blockchain Interpretation} \\
\midrule
Site $(C, J)$ & The ``computational spacetime'' of all possible blockchain evolutions. \\
Object $h$ in $C$ & A specific, valid partial history of the blockchain (a sequence of blocks). \\
Morphism $h \to h'$ & A valid extension of history $h$ to $h'$. \\
Covering Sieve on $h$ & Future extensions of $h$ satisfying a liveness condition (eventual consensus). \\
Presheaf $P$ & Local, potentially inconsistent data indexed by history (e.g., node mempools). \\
Sheaf $S$ & Globally consistent data indexed by history (e.g., the canonical ledger state). \\
Gluing Axiom & The consensus algorithm transforming local views into a global canonical state. \\
Truth Value in $\Omega(h)$ & The set of future histories extending $h$ in which a proposition holds true. \\
\bottomrule
\end{tabular}
\caption{Mapping Topos-Theoretic Concepts to Blockchain Systems}
\label{tab:mapping}
\end{table}

\section{Presheaves of Observation and Sheaves of State}
Having defined the site $(C, J)$, we can now populate this computational spacetime with data. In topos theory, data is represented by presheaves and sheaves \cite{maclane_moerdijk_sheaves}. A presheaf is a collection of data that varies from one context to another, while a sheaf is a presheaf that satisfies an additional consistency condition, allowing local data to be ``glued'' together into a global whole. This distinction provides a powerful formalism for modeling the crucial difference between the chaotic, inconsistent information held by individual nodes and the single, canonical state of the global ledger.

\subsection{Presheaves of Local Views ($P$)}
A presheaf on the category $C$ is formally a contravariant functor $P: C^{\text{op}} \to \mathbf{Set}$ \cite{goguen_unification}. This means it assigns a set $P(h)$ to each object $h \in C$ and a restriction map $P(f): P(h') \to P(h)$ to each morphism $f: h \to h'$. We define the \emph{presheaf of local views}, denoted $P_{\text{view}}$, to capture the raw, uncoordinated state of information across the network.
\begin{itemize}
    \item \textbf{On Objects:} For any history $h \in C$, the set $P_{\text{view}}(h)$ is the set of all possible local states of all nodes in the network whose internal view is consistent with having observed the history $h$. A single element of $P_{\text{view}}(h)$ could represent, for example, the contents of a specific node's memory pool (mempool) of unconfirmed transactions, its set of known candidate chain tips, or its internal database state.
    \item \textbf{On Morphisms (Restriction Maps):} For a morphism $f: h \to h'$, which represents an extension of the chain, the restriction map $P_{\text{view}}(f): P_{\text{view}}(h') \to P_{\text{view}}(h)$ models the act of ``forgetting'' information. It takes a local state consistent with the longer history $h'$ and maps it to the corresponding state at the earlier history $h$. For instance, if a transaction was confirmed in the blocks between $h$ and $h'$, the restriction map would map a local view at $h'$ (where the transaction is confirmed) to a local view at $h$ (where the transaction was still in the mempool).
\end{itemize}
The crucial property of $P_{\text{view}}$ is that it is fundamentally \textbf{inconsistent}. For a given history $h$, the set $P_{\text{view}}(h)$ will contain a multitude of different mempools from different nodes, conflicting candidate blocks, and varied perceptions of the network state. This presheaf accurately models the chaotic, unsynchronized, and informationally fragmented reality of the network's periphery before consensus has been reached. It is a mathematical representation of the ``gossip layer.''

\subsection{The Sheaf of Global State ($S$)}
In contrast to the presheaf of local views, we define the \emph{sheaf of global state}, denoted $S_{\text{state}}$, to represent the single, canonical truth of the ledger.
\begin{itemize}
    \item \textbf{On Objects:} For any history $h \in C$, the set $S_{\text{state}}(h)$ is the \emph{singleton set} containing the single, canonical ledger state (e.g., the complete set of account balances, smart contract storage, and UTXO set) as determined by the deterministic application of the protocol's state transition rules to the sequence of blocks in history $h$.
    \item \textbf{On Morphisms (Restriction Maps):} The restriction maps are defined by the state transition function. For an extension $f: h \to h'$, the state at $h$ is uniquely determined by the state at $h'$ by simply ignoring the state changes effected by the blocks from $h$ to $h'$.
\end{itemize}
$S_{\text{state}}$ is a presheaf by construction, but it possesses a much stronger property: it is a sheaf under the liveness topology $J$.

\subsection{The Sheaf Condition as Consensus}
The connection between the abstract sheaf condition and the concrete process of blockchain consensus is the central thesis of this model. The sheaf axiom provides a formal, mathematical definition of what it means for a system to be able to achieve consensus.
\begin{itemize}
    \item \textbf{Formal Statement of the Gluing Axiom:} A presheaf $S$ is a sheaf if for any object $h$ and any covering sieve $S_{\text{cov}} = \{f_i: h_i \to h\}_{i \in I}$ in $J(h)$, any compatible family of sections $\{s_i \in S(h_i)\}_{i \in I}$ gives rise to a \emph{unique} amalgamated section $s \in S(h)$ that restricts to each $s_i$. A family of sections is ``compatible'' if for any two sections $s_i \in S(h_i)$ and $s_j \in S(h_j)$, they agree on their common ``overlap,'' which means they restrict to the same section on any history $h_k$ that is a predecessor to both $h_i$ and $h_j$.
    \item \textbf{Interpretation as Semantic Unification:} This process can be understood as a form of ``semantic unification'' \cite{goguen_unification}. The compatible local sections $\{s_i\}$ represent locally agreed-upon states in different possible futures $\{h_i\}$. For example, in a fork, $h_1$ and $h_2$ might be two competing one-block extensions of $h$. A compatible family of states would be a state for $h_1$ and a state for $h_2$ that both correctly derive from the same state at $h$. The consensus algorithm is the mechanism that provides the ``gluing'' function. It is a deterministic rule (e.g., the longest-chain rule, BFT finality) that guarantees that from this collection of potential future states, a \emph{unique} canonical state $s$ can be determined for the present moment $h$.
\end{itemize}
The presheaf of local views, $P_{\text{view}}$, is \emph{not} a sheaf. One can easily find a covering sieve of possible futures and compatible local views for which there is no unique amalgamated global view at $h$. The system is in a state of ambiguity. A failure of the sheaf condition for a presheaf signifies that the protocol it models has an unresolvable fork or ambiguity---a consensus failure.

This framework reveals that the entire process of achieving consensus can be modeled formally as the mathematical construction of \emph{sheafification}. Any presheaf $P$ can be transformed into its associated sheaf, denoted $aP$. This process algorithmically resolves the inconsistencies in $P$ to produce a sheaf that best approximates it. The sheafification functor, $a(-)$, models the dynamic process of consensus itself:
\begin{equation}
S_{\text{state}} \cong a(P_{\text{view}})
\end{equation}
Consensus algorithms can thus be rigorously studied by examining the mathematical properties of their corresponding sheafification functors.

\section{The Internal Logic of the Blockchain Topos}
One of the most profound consequences of modeling a system as a topos is that every topos comes equipped with an internal logic \cite{maclane_moerdijk_sheaves}. For a Grothendieck topos of sheaves, this logic is, in general, intuitionistic. This means that certain principles of classical logic, most notably the Law of the Excluded Middle (LEM), which states that for any proposition $\phi$, the statement $\phi \lor \neg\phi$ is true, do not hold universally \cite{goldblatt_topoi, fourman_scott_logic}. This feature provides an exceptionally precise language for reasoning about the epistemological realities of a distributed system, such as uncertainty, finality, and forks.

\subsection{Contextual Truth and the Failure of Excluded Middle}
In the topos of sheaves on the site of computation $(C, J)$, the truth of a proposition is not an absolute, global property. Instead, truth is contextual; a proposition is evaluated \emph{at a specific history} $h \in C$. The failure of LEM is a direct reflection of the system's inherent ambiguity and its dependence on future events. Consider the proposition $\phi \equiv$ ``Transaction $T$ is confirmed.''
\begin{itemize}
    \item \textbf{Truth:} If transaction $T$ is included in a block that is part of the history $h$ (and is sufficiently deep to be considered final), then the proposition $\phi$ is true at stage $h$.
    \item \textbf{Falsity:} If a conflicting transaction (e.g., a double-spend of the same inputs) is confirmed in history $h$, then the protocol dictates that $T$ can never be confirmed. In this case, the proposition $\neg\phi$ is true at stage $h$.
    \item \textbf{Indeterminacy:} Consider the most common case: $T$ has been broadcast to the network and resides in the mempools of many nodes, but it has not yet been included in a block in the history $h$. At this stage, we cannot assert that $\phi$ is true, because network latency, transaction fees, or malicious miners might prevent $T$ from ever being included. Crucially, we also cannot assert that $\neg\phi$ is true, because $T$ might well be included in the very next block.
\end{itemize}
In this indeterminate state, the Law of the Excluded Middle fails; the proposition $\phi \lor \neg\phi$ is not true at stage $h$. The internal logic correctly reflects the operational reality: the truth value of the statement is pending, contingent on the future evolution of the system. Truth is identified with construction or proof, which in this context means inclusion in the canonical chain.

\subsection{Characterizing the Subobject Classifier ($\Omega$)}
In any topos, there exists a special object $\Omega$, called the subobject classifier, which acts as the object of truth values for the internal logic. In a topos of sheaves, $\Omega$ is itself a sheaf, and its sections over an object $h \in C$ are given by the set of all sieves on $h$ that are closed with respect to the topology $J$ \cite{maclane_moerdijk_sheaves}. For simplicity, we can think of $\Omega(h)$ as the set of all sieves on $h$, representing a collection of possible future paths or evolutions of the system.

In the internal logic, the truth value of a proposition $\phi$ at a stage $h$, denoted $[[\phi]]_h$, is a sieve:
\begin{itemize}
    \item A proposition $\phi$ is considered \textbf{true} at stage $h$ (or ``forced'' at $h$) if its truth value $[[\phi]]_h$ is the maximal sieve, which contains all possible future paths.
    \item A proposition $\phi$ is \textbf{false} at stage $h$ if $[[\phi]]_h$ is the empty sieve, meaning there is no possible future in which it becomes true.
    \item If $[[\phi]]_h$ is a non-empty, non-maximal sieve, the proposition is indeterminate.
\end{itemize}

This framework provides a formal language for reasoning about forks and reorganizations. Consider a fork at history $h$ with two competing blocks, $b_1$ and $b_2$. Let $\phi_1$ be the proposition ``block $b_1$ is canonical'' and $\phi_2$ be ``block $b_2$ is canonical.'' At stage $h$, neither $[[\phi_1]]_h$ nor $[[\phi_2]]_h$ is the maximal sieve. They are two distinct, non-trivial sieves representing the two possible branches of the fork. The consensus protocol's fork-choice rule (e.g., the longest-chain rule) is a mechanism for evaluating which of these truth values will eventually be forced toward the maximal sieve.

This perspective recasts security considerations into the language of model theory. A 51\% attack, for instance, is an action by a malicious actor to manipulate the generation of future histories to ensure that the sieve corresponding to their desired version of history ($[[\phi_{\text{malicious}}]]_h$) becomes the maximal sieve. They are not ``breaking'' the logic; they are actively forcing a particular proposition to become true within it, opening up new formal methods for security analysis.

\section{The Yoneda Lemma: The Idealism of the Global View vs. The Reality of the Local Node}
The Yoneda lemma establishes a deep connection between the objects of a category and the functors defined upon it \cite{awodey_category, leinster_category}. In the context of the blockchain topos, this lemma provides a profound statement about the nature of decentralized systems, while highlighting the crucial gap between the abstract, Platonic truth of the mathematical model and the concrete, epistemic reality of a single participant within that system.

\subsection{The Theoretical Power of Yoneda: ``Local Determines Global''}
The Yoneda lemma establishes a natural isomorphism for any object $h$ in a category $C$ and any presheaf $S: C^{\text{op}} \to \mathbf{Set}$:
\begin{equation}
\text{Nat}(\text{Hom}_C(-, h), S) \cong S(h)
\end{equation}
where $\text{Hom}_C(-, h)$ is the representable presheaf mapping an object $x$ to the set of morphisms from $x$ to $h$ \cite{leinster_category}. Let $S$ be the sheaf of global state, $S_{\text{state}}$. The lemma states that the global state at a specific history $h$, $S_{\text{state}}(h)$, is completely determined by the web of all possible ways to probe or view the sheaf from the perspective of all other histories in the category. The object $S_{\text{state}}(h)$ has no hidden internal information; its identity is constituted entirely by its external, relational properties.

This is the mathematical crystallization of the decentralized ethos. The global truth, $S_{\text{state}}(h)$, is not a transcendent fact stored in a special location, but is immanent in the totality of all possible local perspectives and their interrelations.

\subsection{The Practical Pitfall: The Inaccessibility of Counterfactuals}
While the Yoneda lemma provides an ontological truth about the mathematical structure, it does not provide an epistemological or computational path for a participant within the system to arrive at that truth. To compute the global state $S_{\text{state}}(h)$ via this isomorphism, one would need to construct the set of all natural transformations from $\text{Hom}_C(-, h)$ to $S_{\text{state}}$. A single such natural transformation $\alpha$ is a family of functions $\{\alpha_x: \text{Hom}_C(x, h) \to S_{\text{state}}(x)\}$ defined for \emph{every object} $x$ in the category $C$, where each object $x$ is a valid partial history.

A real node participating in the network exists at a single, concrete point within this vast categorical spacetime. It has experienced one specific history, $h_{\text{node}}$, and has direct computational access to its local approximation of $S_{\text{state}}(h_{\text{node}})$. However, it has no computational or epistemic access to the value of $S_{\text{state}}(x)$ for a history $x$ that is \emph{counterfactual} relative to its own experience. These counterfactuals include forks that were abandoned, blocks that were valid but never propagated, or alternative transaction orderings. These are precisely the contextual sections that a real-world node cannot replicate or observe.

The Yoneda lemma describes a ``God's-eye view'' requiring simultaneous knowledge of the entire multiverse of all possible valid histories. A physical node is bound to a single worldline. Its perspective is fundamentally and irreducibly local. Therefore, while the Yoneda lemma guarantees that the global state is theoretically determined by local views in aggregate, it also reveals that no single local participant can ever perform this determination directly.

This tension delineates the boundary between the \emph{semantics} of a distributed protocol and the \emph{algorithm} executed by a single process. The node's algorithm is a practical, heuristic procedure for attempting to align its local state with the true global state $S_{\text{state}}(h)$ defined by the topos. This clarifies the origin of probabilistic finality in protocols like Proof-of-Work: a node can never be 100\% certain its local view corresponds to the true canonical state because it cannot survey the entire space of counterfactual histories. It can only become certain enough as its observed history becomes so computationally dominant that the probability of a deeply buried counterfactual history ever becoming relevant approaches zero.

\section{Conclusions}

This report has developed a rigorous topos-theoretic framework for blockchain architectures, demonstrating that the dynamics of decentralized consensus are natively geometric and logical rather than merely state-transitional. By replacing the classical finite state machine model with a site of computation $(C, J)$, we have successfully reframed the "Problem of Global State" from an ad-hoc implementation challenge into an emergent property of contextual mathematics. Within this computational spacetime, the operational mechanics of distributed systems are unified under a single structural paradigm: 

\begin{itemize}
    \item \textbf{Consensus as Sheafification:} The traditional boundary between fragmented local information (the gossip layer) and the canonical ledger state is formally characterized by the distinction between presheaves and sheaves. Under this view, the execution of a consensus algorithm is mathematically equivalent to the application of a sheafification functor, $a(P_{\text{view}}) \cong S_{\text{state}}$, which systematically maps compatible local views into unique global sections.
    \item \textbf{Intuitionistic Epistemology:} The failure of the Law of the Excluded Middle ($\phi \lor \neg\phi$) within the internal logic of the topos is not an abstract curiosity but a high-fidelity model of network indeterminacy. Truth in a blockchain is fundamentally constructive, requiring explicit block inclusion to function as a categorical proof, while security attacks translate directly into the manipulation of forced truth values within the subobject classifier $\Omega$.
    \item \textbf{The Yoneda Duality:} While the Yoneda lemma provides an ontological guarantee that the global state is fully determined by the aggregate relational structure of local nodes, it simultaneously exposes a critical epistemic boundary. Because physical nodes are bound to a single worldline and cannot survey counterfactual or abandoned histories, they cannot directly compute the global state via natural transformations. This intrinsic inaccessibility of counterfactual sections provides a novel foundation for understanding the necessity of probabilistic finality and heuristic algorithms in real-world systems engineering.
\end{itemize}

To summarize, shifting the analytical paradigm from static state
transitions to the dynamic, geometric process of local alignment
reveals that decentralized truth is inherently immanent,
context-dependent, and actively constructed. The topos $Sh(C, J)$
establishes a foundational semantics for distributed systems, opening
up a promising avenue for the mathematical verification of consensus
resilience, security verification, and the formal model theory of
adversarial computing environments.
\bibliographystyle{plain}
\bibliography{references}

\end{document}